\documentclass[12pt,a4paper]{article} 
\usepackage{graphicx}
\usepackage[latin1]{inputenc} 
\usepackage[T1]{fontenc} 
\usepackage{amsmath}
\usepackage{amsfonts} 
\usepackage{amssymb} 
\usepackage[english]{babel}
\title{On the static Casimir effect with parity-breaking mirrors}
\author{C.~D.~Fosco$^{a}$
and
M.~L.~Remaggi$^{b}$\\
{\normalsize\it $^a$Centro At\'omico Bariloche and Instituto Balseiro}\\
{\normalsize\it Comisi\'on Nacional de Energ\'\i a At\'omica} \\
{\normalsize\it 8400 Bariloche, Argentina.} \\
{\normalsize\it $^b$Facultad de Ciencias Exactas y Naturales}\\
{\normalsize\it Universidad Nacional de Cuyo} \\
{\normalsize\it 5500 Mendoza, Argentina.}} 
\begin{document} 
\date{}
\maketitle
\begin{abstract} 
\noindent We study the Casimir interaction energy due to the vacuum
fluctuations of the Electromagnetic (EM) field in the presence of two
mirrors, described by $2+1$-dimensional, generally nonlocal actions, which
may contain both parity-conserving and parity-breaking terms.  We compare
the results with the ones corresponding to Chern-Simons boundary
conditions, and evaluate the interaction energy for several particular
situations.  
\end{abstract}
\section{Introduction}\label{sec:intro}

The Casimir effect~\cite{booksCasimir} is usually regarded as one of the
most remarkable macroscopic manifestations of the fluctuations (be them
quantum or thermal) of a field when it is subjected to the non trivial
influence of external agents.  The latter usually manifest themselves as
boundary conditions, or as `boundary terms' in the action for the field.
These terms are, by definition, contributions depending only on the field
and its derivatives on the boundary; therefore, they can be interpreted as
due to singular terms (involving generalized functions) in the Lagrangian.
The effect results from the interplay between those external agents
(`mirrors') and the field fluctuations.  In the static version of the
effect, the one we are concerned with here, one considers time-independent
boundary conditions or, equivalently, boundary terms which do not depend
explicitly on time.

A variety of situations can be explored where this effect becomes relevant;
a natural way to exhaust them all, is by either considering fluctuating
fields of different nature for each boundary condition, or by studying the
consequences of imposing different boundary conditions on each given field.
In principle, both the mirrors' geometry and their intrinsic properties are relevant
to the effect.  Having in mind the latter, our aim here is to consider
boundary actions containing both parity-conserving and parity-breaking
terms, for an Abelian gauge field in $3+1$ dimensions~\footnote{`Parity' is
understood here in the $2+1$ dimensional sense, namely, the
reflection along an odd number of spacetime coordinates.}, in the presence
of two zero-width mirrors.
A concrete realization of that kind of boundary term are the effective
actions in $2+1$ dimensions which represent the quantum effects due to a
Dirac field confined to the mirrors' world-volumes, and minimally
coupled to the projection of the gauge field to the world-volume swept by
the boundary.  

Note that the Casimir effect due to Chern-Simons (C-S) boundary conditions
has been studied since the pioneering work of
reference~\cite{Bordag:1999ux}, where it has been shown that the Casimir
force may, for some choices of the parameters, become repulsive.  It is our
aim here to study the problem of including parity-breaking terms in the
{\em boundary action\/}, as opposed to boundary conditions. The two
approaches, although related, are essentially different, a fact that has
been highlighted already in~\cite{Bordag:1999ux}.

There have been several interesting developments related to this kind of
system: the Casimir effect for a spherical region, with C-S like boundary
conditions due to the presence of a $\theta$ term, has been considered
in~\cite{Canfora:2011fd}.  A related trend of research
dealt with the Casimir force for two Chern insulators,
including the full frequency dependence of the conductivity
tensor~\cite{Rodriguez-Lopez:2013pza}. Interesting results have been
obtained also in the context of lattice field
theory~\cite{Pavlovsky:2010zz,Pavlovsky:2010zza}, using numerical
approaches which are naturally formulated within that context.
A noteworthy consequence of having a parity-breaking term manifests itself
even for a single mirror. Indeed, this is the case of the interesting
`quantum Faraday effect' discussed in~\cite{Fialkovsky:2009wm} for the
boundary term due to a massive Dirac fermion in $2+1$ dimensions. 

In spite of the fact, mentioned in~\cite{Bordag:1999ux}, that there is no
exact equivalence between boundary conditions and a boundary action, here
we show how the results presented in~\cite{Bordag:1999ux} can be obtained
by using a judiciously chosen boundary action. As we shall see, it must
contain both parity-breaking and parity-conserving terms. Interestingly,
that is exactly the structure of the leading terms in a small-mass expansion
for the effective action due to a massive Dirac field in $2+1$
dimensions~\cite{Deser:1982vy}.

This paper is organized as follows: in Sect.~\ref{sec:thesys} we define the
system and present the conventions we have adopted to describe it. Its
corresponding Casimir interaction energy is introduced in~\ref{sec:energy}. In
Sect.~\ref{sec:casimir} we consider different particular cases.
The reflection coefficients for either one or two mirrors in the basis of
right and left circularly polarized states is presented in an Appendix. 
In Sect.~\ref{sec:conc} we present our conclusions.

\section{The system}\label{sec:thesys}
Within the functional integral formalism, that we shall adopt here, it is
convenient to define the system in terms of its Euclidean action ${\mathcal
S}(A)$, for the Abelian gauge field $A_\mu$. We assume ${\mathcal
S}(A)$ to have the following structure:
\begin{equation}
	{\mathcal S}(A) \;=\; {\mathcal S}_0(A) \,+\, {\mathcal S}_I(A) \;,
\end{equation}
where ${\mathcal S}_0(A)$ denotes the free EM action: 
\begin{equation}
	{\mathcal S}_0(A) \;=\; \frac{1}{4} F_{\mu\nu} F_{\mu\nu} \;,\;\;
	F_{\mu\nu}
= \partial_\mu A_\nu - \partial_\nu A_\mu\;,\;\; \mu = 0,\,1,\,2,\,3 \;,
\end{equation}
and ${\mathcal S}_I$ represents the coupling between the field and the
mirrors.

We assume, for the time being, that there are just two flat infinite
mirrors, located at $x_3=0$ and $x_3=a$, and denoted by $L$ and $R$,
respectively.  Since the spatial region occupied by each mirror is a plane,
one may interpret ${\mathcal S}_I$ as defining two $2+1$-dimensional field
theories, involving the components of the gauge field projected to the
corresponding reduced spacetime.  We recall that, in the case of {\em
perfectly conducting mirrors}, the role of those $2+1$ dimensional theories
is tantamount to imposing the vanishing of the components of the electric
field which are parallel to the mirrors, as well as the component of the
magnetic field which is normal to them.  This can be achieved, for example,
by introducing appropriate auxiliary fields which implement those
conditions, or by taking the proper limit from certain actions
corresponding to imperfect mirrors~\cite{Fosco:2012gp}. 

In this article, we shall consider a rather general case, obtained by
assuming that the  corresponding localized actions are quadratic and gauge
invariant, but we allow for the existence of both parity-conserving and
parity-breaking terms.
More explicitly, the form of ${\mathcal S}_I$ is:
\begin{equation}
 {\mathcal S}_I \;=\; {\mathcal S}^{(L)} \,+\, {\mathcal S}^{(R)} \;,
 \end{equation}
 where  ${\mathcal S}^{(L,R)}$ denotes the action concentrated on the
 mirror at $x_3=0,\,a$, respectively. Each one of these terms may contain
 both parity-even ($e$) and parity odd ($o$) terms. It is convenient to
 introduce a special notation for the parallel coordinates (including the
 time $x_0$): $x_\parallel = (x_\alpha)$, where indices from the beginning
 of the Greek alphabet will be assumed to run over the values $0,\,1,\,2$.
 Then, we may write formally ${\mathcal S}^{(L)}$, say, as follows:
 \begin{align}
& {\mathcal S}^{(L)} \;=\;  {\mathcal S}_e^{(L)} \,+\, {\mathcal S}_o^{(L)}
 \nonumber\\
& {\mathcal S}_e^{(L)} \;=\; \int d^4x \, \delta(x_3)  \frac{1}{4}
 F_{\alpha\beta}\, f^{(L)}_e(-\partial_\parallel^2) \, F_{\alpha\beta}
 \nonumber\\
& {\mathcal S}_o^{(L)} \;=\; \,\int d^4x \,
 \delta(x_3) \, \frac{i}{2} \, \varepsilon_{\alpha\beta\gamma} A_\alpha \,
 f^{(L)}_o(-\partial_\parallel^2) \, \partial_\beta A_\gamma  \;,
\end{align}
where $\varepsilon_{\alpha\beta\gamma}$ denotes the Levi-Civita symbol in
$2+1$ dimensions, and $f^{(L)}_{e,o}$ have been written as functions of
$-\partial_\parallel^2$ in order to indicate that they will be, in general,
nonlocal kernels in coordinate space.  For the $R$ mirror, the structure is
quite similar; the relevant changes are that the $\delta$-function must by
shifted: \mbox{$\delta(x_3) \to \delta(x_3-a)$} and, since the mirrors will
not be regarded as necessarily identical in their properties, the kernels
may be different.  Thus, in ${\mathcal S}^{(R)}$ one also has to make the
replacement: \mbox{$f^{(L)}_{e,o} \to  f^{(R)}_{e,o}$}. Note that the
kernels will have the mass dimensions: $[f^{(L,R)}_e] = -1$ and 
$[f^{(L,R)}_o] = 0$.

Assuming, however, that the mirrors' properties are translation invariant
and time independent (i.e., invariant under translations in the $x_\parallel$
coordinates), they will be local in momentum space.
Note that the cases of perfect mirrors, or mirrors described purely by
a C-S term may be obtained by taking particular limits for the
kernels.

We see that, introducing Fourier transformations with respect to the
parallel coordinates:
\begin{equation}
	\widetilde{A}_\alpha(k_\parallel,x_3) \;=\; \int d^3x_\parallel \,
	e^{- i k_\parallel \cdot x_\parallel} \,A_\alpha(x_\parallel,x_3) \;. 
\end{equation}
we may write: 
\begin{align}
& {\mathcal S}_e^{(L)} \;=\; \frac{1}{2} \int \,
	\frac{d^3k_\parallel}{(2\pi)^3} \,
	\widetilde{A}^*_\alpha(k_\parallel,0)  \; \alpha^{(L)}_P(k_\parallel) 
	\; P_{\alpha\beta}(k_\parallel) \; 
	\widetilde{A}_\beta(k_\parallel,0) \nonumber\\
& {\mathcal S}_o^{(L)} \;=\; \frac{1}{2} \, \int \,
	\frac{d^3k_\parallel}{(2\pi)^3} \,
	\widetilde{A}^*_\alpha(k_\parallel,0) \; \alpha^{(L)}_Q(k_\parallel)
	\; Q_{\alpha\beta}(k_\parallel) \; 
	\widetilde{A}_\beta(k_\parallel,0)
\end{align}
where 
\begin{equation}\label{eq:aux1}
\alpha^{(L)}_P(k_\parallel) \,\equiv\, k_\parallel^2 \,
f^{(L)}_e(k_\parallel^2)  \;,\;\; 
\alpha^{(L)}_Q(k_\parallel) \,\equiv\, -
|k_\parallel| \, f^{(L)}_o(k_\parallel^2) \;,
\end{equation}
and we have introduced:
\begin{equation}
 P_{\alpha\beta}(k_\parallel) \;=\; \delta_{\alpha\beta} - \frac{k_\alpha
 k_\beta}{k_\parallel^2} \;\;,\;\;\;
 Q_{\alpha\beta}(k_\parallel) \;=\; \varepsilon_{\alpha\gamma\beta}
	\frac{k_\gamma}{|k_\parallel|} \;. 
\end{equation}
These tensors satisfy algebraic relations which, using a matrix notation, adopt the
form:
\begin{equation}
P^2 \,=\, P \;,\;\;\; Q^2 \,=\, -P \;,\;\;\; P Q \,=\, Q P \,= Q \;.
\end{equation}
To simplify our next developments, it is convenient to have a complete set
of orthogonal projectors for the space of $3 \times 3$ Hermitian matrices,
which naturally arise in the Fourier representation. The orthogonality
property allows one to deal with each invariant subspace separately,
naturally decomposing the original problem a set of one-dimensional decoupled
problems. 

Those projectors can
be built by inspection, taking into account the relations above.
Indeed, defining $P^{\pm} \equiv \frac{P \pm i Q}{2}$ and  $P' \equiv I -
P$ ($I$ denotes the identy matrix), we see that:
\begin{align}
& P^+ + P^- + P' \,=\, I \;,\;\;  (P^{\pm})^2 \,=\,P^{\pm} \;,\;\;P'^2
\,=\,P'\;, \nonumber\\
& P^+ P^- \,=\,P^- P^+ \,=\,P^{\pm} P' \,=\,P' P^{\pm} \,=\, 0 \;.       
\end{align}

Then, using the Fourier representation above, we have for the full action
${\mathcal S}$ (in the Feynman gauge) the following expression:
\begin{align}\label{eq:fours}
& {\mathcal S}(A) \;=\; \frac{1}{2} \,\int \frac{d^3k_\parallel}{(2\pi)^3}
\, \int dx_3 \; \Big\{ \tilde{A}^*_3(k_\parallel,x_3) \; (-\partial_3^2 +
k_\parallel^2) \; \tilde{A}_3(k_\parallel,x_3) \nonumber\\ 
& + \; \tilde{A}^*_\alpha(k_\parallel,x_3) \; (-\partial_3^2 +
k_\parallel^2) \; \tilde{A}_\alpha(k_\parallel,x_3) \nonumber\\ 
& +\;\tilde{A}^*_\alpha(k_\parallel,x_3) \; \delta(x_3) \; 
\big[ \alpha^{(L)}_- (k_\parallel)  P^+_{\alpha\beta}(k_\parallel)   
\,+\, \alpha^{(L)}_+ (k_\parallel)  P^-_{\alpha\beta}(k_\parallel) \big] 
\tilde{A}_\beta(k_\parallel,x_3) \nonumber\\   
& +\;\tilde{A}^*_\alpha(k_\parallel,x_3) \; \delta(x_3-a) \;  
\big[ \alpha^{(R)}_- (k_\parallel)  P^+_{\alpha\beta}(k_\parallel)   
\,+\, \alpha^{(R)}_+ (k_\parallel)  P^-_{\alpha\beta}(k_\parallel) 
\big] \tilde{A}_\beta(k_\parallel,x_3)  \Big\} \;, 
\end{align}
with 
\begin{equation}\label{eq:aux2}
\alpha^{(L,R)}_\pm \,=\, \alpha^{(L,R)}_P \pm i  \alpha^{(L,R)}_Q \;.
\end{equation}

\section{The interaction energy}\label{sec:energy}
The vacuum energy $E$ of the EM field in the presence of the mirrors, may
be written in terms of the Euclidean vacuum transition amplitude ${\mathcal
Z}$ for a time evolution of length $T$:  
\begin{equation}
E \;=\; - \lim_{T \to \infty} \Big( \frac{1}{T} \, \log {\mathcal Z} \Big)
\;,
\end{equation}
where ${\mathcal Z}$ can be represented as the functional integral:
\begin{equation}
{\mathcal Z} \;=\; \int {\mathcal D}A \; e^{- {\mathcal S}(A)}\;.
\end{equation} 
Translation invariance along the parallel coordinates suggests to use the
Fourier transformation implemented in (\ref{eq:fours}) in order to evaluate
the functional integral.  Besides,  the Fourier transformed
gauge field may be decomposed, for each set of values of $x_3$ and
$k_\parallel$ in terms of four orthonormal unit vectors, which we will
denote by: $\hat{e}^{(+)}$,  $\hat{e}^{(-)}$, $\hat{e}^{(k)}$, and
$\hat{e}^{(3)}$.
The $\hat{e}^{(3)}$ vector is parallel to the $x_3$ axis, i.e., its
$\mu$ component is $\hat{e}^{(3)}_\mu = \delta^3_\mu$. The other three
vectors are in the orthogonal subspace to the one generated by
$\hat{e}^{(3)}$; one of them, $\hat{e}^{(k)}$, points along $k_\parallel$,
while $\hat{e}^{(\pm)} \equiv \frac{\hat{e}^{(1)} \pm i
\hat{e}^{(2)}}{\sqrt{2}}$, with $\hat{e}^{(1)}$ and $\hat{e}^{(2)}$, 
orthogonal to $k_\parallel$, are such that $\hat{e}^{(1)}$, $\hat{e}^{(2)}$
and $\hat{e}^{(k)}$ form a right-handed orthogonal triplet. 

Thus, we may decompose $\tilde{A}_\mu(k_\parallel,x_3)$ as follows:
\begin{align}
	\tilde{A}_\mu(k_\parallel,x_3)\;= &\; C_+(k_\parallel,x_3)
	\hat{e}^{(+)}_\mu \;+\;  C_-(k_\parallel,x_3) \hat{e}^{(-)}_\mu
	\nonumber\\
	& +\;  C_k(k_\parallel,x_3) \hat{e}^{(k)}_\mu \;+\;
	C_3(x_\parallel,x_3) \hat{e}^{(3)}_\mu \;.
\end{align}
The Fourier transformed action then becomes (we omit the arguments in the
coefficients $C$, for the sake of clarity):
\begin{align}\label{eq:fours1}
	{\mathcal S}(A) =& \frac{1}{2}\int \frac{d^3k_\parallel}{(2\pi)^3}
\int dx_3 \Big\{ 
C^*_+\big[-\partial_3^2 +
k_\parallel^2 \,+\, \delta(x_3) \, \alpha^{(L)}_- (k_\parallel) 
+ \delta(x_3-a)\alpha^{(R)}_- (k_\parallel) \big] C_+ 
\nonumber\\   
& + \;  C^*_- \; \big[-\partial_3^2 +
k_\parallel^2 \,+\, \delta(x_3) \,\alpha^{(L)}_+ (k_\parallel) 
+ \delta(x_3-a) \,\alpha^{(R)}_+ (k_\parallel) \big] C_- 
\nonumber\\   
& + \;  C^*_k \; (-\partial_3^2 +
k_\parallel^2 ) C_k \,+\, C^*_3 \; (-\partial_3^2 +
k_\parallel^2) \; C_3 \Big\} \;. 
\end{align}
The action thus becomes the sum for each $k_\parallel$, of four independent
actions, each one corresponding to a single degree of freedom, represented
by the corresponding coefficient $C$. The functional integration measure
factorizes with respect to $k_\parallel$ (each value can be treated
separately) and also, for each $k_\parallel$, into the product the measures
for each coefficient.

Since $C_k$ and $C_3$ do not see the
mirrors, they can be discarded when evaluating the effect of the mirrors on
the vacuum energy. Taking also into account that log of
${\mathcal Z}$ becomes extensive in $T$ and in the area $L^2$ of the
mirrors,  the energy per unit area, ${\mathcal E}$, becomes:
\begin{equation}
{\mathcal E} \;=\; -\int \frac{d^3k_\parallel}{(2\pi)^3} \,  \log [{\mathcal
Z}_{k_\parallel}] \;,
\end{equation}
where 
\begin{equation}
{\mathcal Z}_{k_\parallel} \;=\; 
{\mathcal Z}_{k_\parallel}^{(+)} \,{\mathcal Z}_{k_\parallel}^{(-)} \;, 
\end{equation}
where each factor above corresponds to the functional integral over the
respective coefficient, and may therefore be expressed formally as a
functional determinant:
\begin{equation}
	{\mathcal Z}_{k_\parallel}^{(\pm)}\;=\; 
	\Big( \det\big[-\partial_3^2 + k_\parallel^2 \,+\,
		\delta(x_3) \, \alpha^{(L)}_{\mp} (k_\parallel) + \delta(x_3-a)
	\alpha^{(R)}_{\mp} (k_\parallel) \big] \Big)^{-\frac{1}{2}}\;.  
\end{equation}

Finally, taking into account the known results about the functional
determinants of the kind arising in the equation
above~\cite{CcapaTtira:2011ga}, we see that the
energy per unit area may be written as follows:
\begin{equation}\label{eq:end}
	{\mathcal E} \;=\; \frac{1}{2}\int \frac{d^3k_\parallel}{(2\pi)^3} \,  
	\log\Big[ \big( 1 -  r_-^{(L)} r_-^{(R)} e^{- 2 |k_\parallel| a }\big)  
	\big( 1 -  r_+^{(L)} r_+^{(R)} e^{- 2 |k_\parallel| a }\big)  \Big]\;.
\end{equation}
We have introduced:
\begin{equation}
	r_{\pm}^{(L,R)} \;=\; \frac{\alpha_\pm^{(L,R)}}{\alpha_\pm^{(L,R)}+
	2 |k_\parallel|} \;,
\end{equation}
which play the role of Euclidean reflection coefficients.

It is interesting to note that the energy of the system may be thought of
as decoupled between two contributions, each one corresponding to either left
or right circular polarization modes. Based on these modes, a useful
parametrization of the reflection coefficients (inspired
by~\cite{Bordag:1999ux}) is the following:
\begin{equation}
r_+^{(L)} \;=\;  - |r^{(L)}|  e^{2 i \delta_L} \;, 
r_+^{(R)} \;=\;  - |r^{(R)}|  e^{- 2 i \delta_R}  
\end{equation}
(the minus signs  amount to a phase convention for $\delta_{L,R}$).
This allows us to write for the energy:
\begin{equation}\label{eq:enda}
	{\mathcal E} \;=\; \frac{1}{2}\int \frac{d^3k_\parallel}{(2\pi)^3} \,  
	\log\Big( 1 -   2 \,|r^{(L)}| |r^{(R)}|  \rm{cos} (2\delta) \,  e^{- 2
	|k_\parallel| a }  +  |r^{(L)}|^2 |r^{(R)}|^2 e^{- 4 |k_\parallel|
	a }\Big)\;,
\end{equation}
where $\delta = \delta_L - \delta_R$.

\section{Results and discussion}\label{sec:casimir}

Let us first show how one can recover the result of imposing C-S
boundary conditions, considered in~\cite{Bordag:1999ux}.
That situation involves both parity breaking and parity conserving boundary
terms, since the boundary conditions mix the parallel components of the
electric field with the parallel components of the magnetic field, and the
normal component of the magnetic field with the normal component of the
electric field. By inspection of the boundary conditions due to the
boundary action we consider in this article, recalling (\ref{eq:aux1}),
(\ref{eq:aux2}),  we chose:
\begin{align}
& f_e^{(L)} \;=\;f_e^{(R)} \,=\, - \frac{1}{|k_\parallel|} \nonumber\\
& f_o^{(L)} \;=\;  \theta(0) \;,\;\; f_o^{(R)} \;=\;  - \theta(a) \;,
\end{align}
where the minus sign in the last equation is just to be consistent with the
choice made in~\cite{Bordag:1999ux} to introduce the boundary conditions
(namely, the normals corresponding to the two surfaces are opposite).
Thus, 
\begin{equation}
	r_L \;=\; - \frac{1 + i \theta(0)}{1- i \theta(0)}\;,\;\;  
	r_R \;=\; - \frac{1 - i \theta(a)}{1+ i \theta(a)} \;. 
\end{equation}
Both have modulus equal to one, and are therefore pure phases. Defining:
\begin{equation}
r_L \;=\; - e^{ 2 i \delta_0 } \;,\;\; r_R \;=\; - e^{- 2 i \delta_a }
\end{equation}
with $\delta_{0,a} \equiv \arctan \theta(0,a)$, we see that the expression
(\ref{eq:end}) becomes:
\begin{eqnarray}\label{eq:endeb}
	{\mathcal E} &=& \frac{\varphi_b(\delta)}{a^3} \nonumber\\ 
	{\varphi_b(\delta)} &=& \frac{1}{32 \pi^2}\int_0^\infty dk \,k^2 \,  
	\log\Big( 1 -   2 \, \rm{cos}(2\delta) \,  e^{-k}  +   e^{- 2 k}\Big)\;,
\end{eqnarray}
with $\delta = \delta_0 - \delta_a = {\rm arctan}\big(\frac{\theta(0) -
\theta(a)}{ 1 + \theta(0) \theta(a)}\big)$, which agrees with the result
in~\cite{Bordag:1999ux}.

It is worth noting that this choice of boundary term can also be understood
as the most general one such that there are no dimensionful constants in
its kernels. Indeed, coming back to the form of the boundary action, we see
that it can be written (for the $L$ mirror, say) as follows:
\begin{equation}
	S^{(L)} \;=\; \int d^4x \delta(x_3) \,
	\big( \frac{1}{4} F_{\alpha\beta}
	(-\partial_\parallel^2)^{-1/2} F_{\alpha\beta}
		\,+\,
			\frac{i\theta_L}{2} \varepsilon_{\alpha\beta\gamma}
			A_\alpha \partial_\beta A_\gamma \big) \;. 
\end{equation}
It is worth noting that essentially the same structure arises as the
one-loop effective action for a massless Dirac fermion, with the parity-odd
term reflecting the existence of the parity anomaly.

We can also consider boundary terms which only contain parity breaking
terms such that the violation of parity is maximal.  This case amounts to
taking $f_e^{(L,R)}=0$, and $f_o^{(L,R)}=
\theta_{L,R}$ where each $\theta_{L,R}$ is a dimensionless constant.

The result for ${\mathcal E}$ in this case, may be put as follows:
\begin{equation}
{\mathcal E}\;=\; \frac{\varphi(\theta_L, \theta_R)}{a^3}
\end{equation}
with the dimensionless function $\varphi$:
\begin{align}
	\varphi(\theta_L, \theta_R) \,=\,\frac{1}{32\pi^2} \, \int_0^\infty
	dk k^2 \, \log\Big[ 1 & + \frac{ \theta_L^2 \theta_R^2}{( 4 +
\theta_L^2) (4 + \theta_R^2)} e^{- 2 k} \nonumber\\
& +\, \frac{8\theta_L \theta_R - 2 \theta_L^2 \theta_R^2}{( 4 +
\theta_L^2) (4 + \theta_R^2)} e^{-k} \Big] \;.
\end{align}
In particular, for identical mirrors: $\theta_L = \theta_R \equiv \theta$,
\begin{equation}
	\varphi(\theta, \theta) \,\equiv \, \varphi_g(\theta) \;=\;
	\frac{1}{32\pi^2} \, \int_0^\infty
	dk k^2 \, \log\Big[ 1  - \frac{\theta^2}{( 4 + \theta^2)^2} e^{-k}
	[\theta^2 (2 - e^{-k}) - 8 ] \Big] \;.
\end{equation}
On the other hand, if the C-S coefficients have equal modulus and
opposite signs: $\theta_L = -\theta_R \equiv \theta$,  
\begin{equation}
	\varphi(\theta,-\theta) \,\equiv \, \varphi_u(\theta) \;=\;
	\frac{1}{32\pi^2} \, \int_0^\infty
	dk k^2 \, \log\Big[ 1  - \frac{\theta^2}{( 4 + \theta^2)^2} e^{-k}
	[\theta^2 (2 - e^{-k}) + 8 ] \Big] \;.
\end{equation}

Finally, the result corresponding to a perfect mirror (L) facing a C-S mirror (R) with
constant $\theta$ may be obtained by evaluating the general expression for
the interaction energy for the case $f^{(L)}_o \equiv 0$, $f^{(L)}_e \to
\infty$, and  $f^{(R)}_e \equiv 0$, $f^{(L)}_o \equiv \theta$.

The resulting expression may be put in the form:
${\mathcal E} \;=\; \frac{1}{a^3} \varphi_c(\theta)$, with 
\begin{equation}
\varphi_c(\theta) \,=\,\frac{1}{32\pi^2} \, \int_0^\infty
dk k^2 \, \log\Big[ 1  - \frac{\theta^2}{4 + \theta^2} e^{- k}
(2 - e^{-k}) \Big] \;.
\end{equation}

\begin{figure}[h!]
\includegraphics[width=\linewidth]{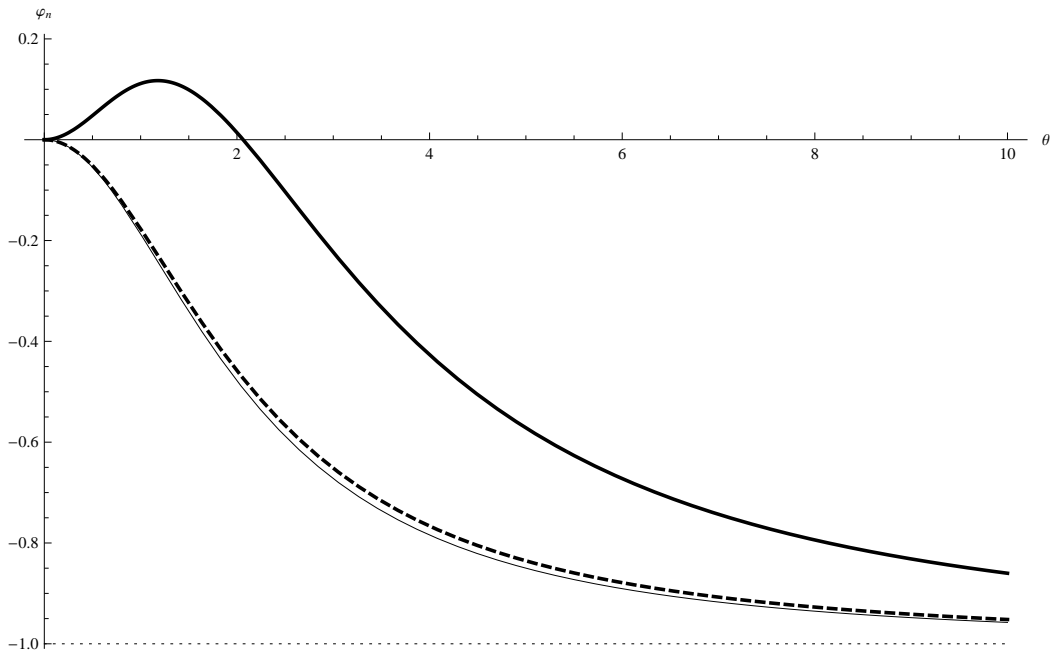}
\caption{$\varphi_n$ vs. $\theta$ for: (a) Two perfect conductors (dotted
	line), (b) two identical C-S mirrors (dashed line), (c)two C-S mirrors with
	opposite coefficients (thin line), (d) a perfect conductor facing a C-S
mirror (thick line).}
\label{fig1}
\end{figure}

In order to have a qualitative idea of the behaviour of the energy for the
different cases we have considered before, we first note that all of them
have the same dependence with the distance (since there is no dimensional
constant in the problem). They have therefore the structure ${\mathcal E} =
\frac{\varphi(\theta)}{a^2}$, with a $\varphi$ which may be $\varphi_g$,
$\varphi_u$ or $\varphi_c$, depending on the case considered.
The same happens for two perfect conductors, for which we recover the
well-known result, given by ${\mathcal E} = \frac{\varphi_p}{a^3}$, with
$\varphi_p = -\frac{\pi^2}{720}$. Using this constant as
reference, in Fig.~\ref{fig1}
we plot a normalized version of $\varphi$ for each case, namely, $\varphi_n
\equiv \frac{\varphi}{\frac{\pi^2}{720}}$, as a function of the C-S
coefficient $\theta$ for the particular cases considered before. The dotted
horizontal line represents the case of two perfect conductors as a
reference, the dashed line corresponds to two identical C-S mirrors, the
solid thin line
corresponds to two C-S mirrors with opposite sign and equal modulus
coefficients and the solid thick line represents a perfect conductor facing
a C-S mirror.

Note that, in all C-S cases, the energy tends to the one of two perfect
conductors as $\theta$ tends to infinity, and their values have no
significant difference already after $\theta=50$. 

Most notably, for the case of two purely C-S mirrors with equal
coefficients, we see that the constant becomes negative when $\theta$ is
lower than $\sim 2.07$, which corresponds to a repulsive
force between the mirrors. It exhibits a non monotonous behaviour as a
function of $\theta$, what may be understood as a consequence of the fact
that for  $\theta$ tending to zero, one should expect the energy to vanish. 
The existence of another zero at approximately $\theta = 2.07$ implies the
non-monotonous character of the coefficient. 

\section{Conclusions}\label{sec:conc}
We have obtained a general expression for the Casimir energy corresponding
to two mirrors, describing matter which may contain both parity-conserving
and parity-breaking terms. Based on the general result, we have shown that
the results obtained by introducing a boundary condition (rather than
boundary action) may be recovered as the result of using a special class of
boundary terms. These should involve no dimensionful constant in their
definition. As such, they have a very similar structure to the one that one
would obtain as the effective action due to a massless Dirac field in $2+1$
dimensions.

The general expression that one has for the energy per unit area in the
general case, where there may exist both parity-conserving and
parity-breaking terms~\ref{eq:enda}, shows that only in the case
$\delta =0$ the energy becomes the sum of two equal contributions. In other
words, the contribution to the vacuum energy of each (left-handed and
right-hand) circular polarization mode is the same only when the phases of the
reflection coefficients are equal. We have also checked, at the level of
the reflection coefficients, that when the relative phase $\delta$
vanishes, the reflection coefficients (see Appendix) become equal for both
handednesses, both for one or two mirrors.

In the particular examples we have considered, we have found and
interpreted an interesting phenomenon when parity violation is maximal
(i.e., no parity-conserving term), namely, the existence of a non
monotonous behaviour of the energy as
a function of the strength of the C-S terms, when assumed to be equal.     
\section*{Acknowledgements}
This work was supported by ANPCyT, CONICET, UBA and UNCuyo.
\newpage
\section*{Appendix: Reflection coefficients}
In order to relate the functions and parameters used in the class of models
considered to more directly observable magnitudes, we calculate here the
reflection coefficients for either one or two mirrors.

Reflection coefficients are relevant to a scattering situation, therefore
it is rather natural to use here the real-time formalism. Thus we assume in
what follows the continuation back to real-time of the corresponding
Euclidean objects has been implemented.  
\subsection{One mirror}
Let us first consider the case of only one mirror, located at $x^3=0$. The
classical equation of motion in this case, assuming the Feynman gauge
is used, is given by:
\begin{equation}
	\Box A^\mu(x) \;+\; \delta(x^3) \, g^{\mu\alpha} 
	\, \Pi^{(L)}_{\alpha\beta} \, g^{\beta\mu} \, A_\nu(x)
	\;=\; 0 \;,
\end{equation}
with 
\begin{equation}
	\Pi^{(L)}_{\alpha\beta} \;\equiv\; f_e^{(L)}(\partial_\parallel^2) \,
	(\partial_\parallel^2
	g_{\alpha\beta} - \partial_\alpha \partial_\beta ) \,+\,
	f_o^{(L)}(\partial_\parallel^2) \,
	\epsilon_{\alpha\gamma\beta} \partial^\gamma \;
\end{equation}
(where $g_{\mu\nu} = g^{\mu\nu} = {\rm diag}(1,-1,-1,-1)$ and
$\partial_\parallel^2 \equiv \partial_\alpha \partial^\alpha$).

One can solve the equation above with scattering boundary conditions.
We propose a normally incident wave with wave vector $k^3 =
+\sqrt{k_\parallel^2} \equiv k$, and we still have the freedom of fixing its
polarization (two independent components). It may be seen that the
left and right circular polarization vectors diagonalizes the problem in
the sense that the corresponding scattering matrices do not mix. 
Thus, we consider incident waves $\tilde{A}_I^\mu(x^3)$ such that
$\tilde{A}_I^3 = 0$, and:
\begin{equation}
	\tilde{A}_I^\alpha(x^3)\,=\, \epsilon_{\pm}^\alpha \, e^{i k x^3} \;.
\end{equation}
We see that, for $x^3 < 0$, the full solution becomes:
\begin{equation}
\tilde{A}^\alpha(x^3) \;=\; \epsilon_{\pm}^\alpha \, e^{i k x^3} 
\,+\, r_\pm \, \epsilon_{\pm}^\alpha \, e^{-i k x^3} 
\end{equation}
where the reflection coefficient $r_\pm$, which determines the reflected
wave for each polarization, is  given by~\cite{Fialkovsky:2009wm}:
\begin{equation}
	r_\pm \;=\; \frac{- i \alpha^{\mp}}{ 2 k + i \alpha^\mp} \;,
\end{equation}
where $\alpha^\mp$ are the real-time counterparts  of the homonymous
objects introduced in the Euclidean formalism; namely:
\begin{equation}
\alpha^\pm \;=\; -k^2 f_e \pm f_o \;.
\end{equation}
\subsection{Two mirrors}
The system consists now of the two mirrors. We see that the reflection
coefficients for this case are also diagonal in the circular polarization
basis. They may be written as follows:
\begin{equation}
r_\pm \,=\, \frac{1}{2 i k} \frac{ \alpha^\mp_{(L)} + \alpha^\mp_{(R)} e^{2 i
k a} + \frac{i}{2 k} \alpha^\mp_{(L)} \alpha^\mp_{(R)} (1  -  e^{2 i k a})}{1 + \frac{i}{2 k}
(\alpha^\mp_{(L)} + \alpha^\mp_{(R)}) - \frac{1}{(2 k)^2} \alpha^\mp_{(L)} \alpha^\mp_{(R)}
(1 - e^{2 i k a})} 
\end{equation}

\end{document}